\documentclass[  reprint,  showpacs,  onecolumn, secnumarabic, amsmath,
amssymb,  aps, pra, nobibnotes, floats, superscriptaddress]{revtex4}%
\usepackage{amsmath}
\usepackage{graphicx}
\usepackage{amsfonts}
\usepackage{amssymb}%
\providecommand{\U}[1]{\protect\rule{.1in}{.1in}}
\providecommand{\U}[1]{\protect\rule{.1in}{.1in}}
\newtheorem{theorem}{Theorem}
\newtheorem{acknowledgement}[theorem]{Acknowledgement}

\begin{document}
\title{{\large Electronic Structure of Super Heavy Atoms. Revisited.}}
\author{D. M. Gitman}
\email{gitman@if.usp.br}
\affiliation{Institute of Physics, University of S\~ao Paulo, Brazil}
\author{A. D. Levin}
\affiliation{Institute of Physics, University of S\~ao Paulo, Brazil}
\author{I.V. Tyutin}
\email{tyutin@lpi.ru}
\author{B.L. Voronov}
\affiliation{I.E. Tamm Theory Department, P.N. Lebedev Physical Institute, Russia}

\begin{abstract}
The electronic structure of an atom with $Z\leq Z_{\mathrm{c}}=137$ can be
described by the Dirac equation with the Coulomb field of a point charge $Ze$.
It was believed that the Dirac equation with $Z>Z_{\mathrm{c}}$ meets
difficulties because the formula for the lower energy level of the Dirac
Hamiltonian formally gives imaginary eigenvalues. But a strict mathematical
consideration shows that difficulties with the electronic spectrum for
$Z>Z_{\mathrm{c}}$ do not arise if the Dirac Hamiltonian is correctly defined
as a self-adjoint operator. In this article, we briefly summarize the main
physical results of that consideration in a form suitable for physicists with
some additional new details and numerical calculations of the electronic spectra.
\end{abstract}
\pacs{03.65.Pm, 31.10.+z}

\maketitle

\section{Introduction\label{S1}}

The question of electronic structure of an atom with large charge
number\textrm{ }$Z$ of the nucleus, especially with $Z$ that is more than the
critical value $Z_{\mathrm{c}}=\alpha^{-1}\simeq137,04$, where $\alpha$ is the
finite structure constant, is of fundamental importance. The formulation of
QED cannot be considered really completed until an exhaustive answer to this
question is given. Although nuclei with overcritical charges can hardly be
synthesized (at present, the maximum is $Z=118$), the existing heavy nuclei
can imitate the supercritical Coulomb fields at collision. Nuclear forces can
hold the colliding nuclei together for $10^{-19}s$ or more. This time is
enough to effectively reproduce the experimental situation where the electron
experiences the supercritical Coulomb field~\cite{GreMuR85}.

The electronic structure of an atom with $Z\leq Z_{\mathrm{c}}$ can be
described by the Dirac equation, which gives relativistic electronic spectra
in agreement with experiment~\cite{BetSa57}. For such $Z$ a complete set of
solutions of the Dirac equation exists; the corresponding Coulomb field does
not violate the vacuum stability, therefore, the Furry picture in QED can be
constructed, and the relativistic quantum mechanics of an electron in such a
Coulomb field based on the Dirac equation is self-consistent. It was believed
that the Dirac equation with $Z>Z_{\mathrm{c}}$ meets difficulties
\cite{Dirac28,Rose61,AkhBe69}. One of the standard arguments is that the
formula for the lower energy level,
\begin{equation}
E=mc^{2}\sqrt{1-\left(  Z\alpha\right)  ^{2}}\,, \label{1}%
\end{equation}
formally gives imaginary result for $Z>Z_{\mathrm{c}}$. This difficulty of the
imaginary spectrum was attributed to an inadmissible singularity of the
supercritical Coulomb field at the origin for a relativistic electron, see
\cite{ZelPo72}. It was believed that this difficulty can be eliminated if a
nucleus of some finite radius $R$ is considered. It was shown that with
cutting off the Coulomb potential with $Z<173$ at a radius $R\sim
1,2\times10^{-12}\mathrm{cm}$, the Dirac Hamiltonian has physically meaningful
spectrum and eigenstates \cite{PomSm45,Popov}. But even in the presence of the
cutoff, another difficulty arises at $Z\sim173$. Namely, the lower bound state
energy descends to the upper boundary $E=-mc^{2}$ of the lower continuum, and
it is generally agreed that in such a situation, the problem can no longer be
considered a one-particle one because of the electron-positron pair
production, which, in particular, results in a screening of the Coulomb
potential of the nucleus. Probabilities of the particle production in the
heavy-ion collisions were calculated within the framework of this
conception~\cite{GreMuR85}. Unfortunately, experimental conditions for
verifying the corresponding predictions are unavailable at present.

Not disputing the fact that taking account of a finite size of the nucleus
corresponds to a more realistic setting up the problem, we do not agree with
the assertion that the Dirac Hamiltonian with the Coulomb field of
overcritical point-like nucleus charge is inconsistent. The above-mentioned
difficulties with the spectrum for $Z>Z_{\mathrm{c}}$ do not arise if the
Dirac Hamiltonian is correctly defined as a self-adjoint (s.a.) operator. A
first heuristic attempt in this direction is due to \cite{Case50}. A rigorous
mathematical treatment of all the aspects of this problem including a spectral
analysis of the Hamiltonian based on the theory of s.a. extensions of
symmetric operators and the Krein method of guiding functionals was presented
in \cite{VorGiT07a,VorGiT,book}. It was demonstrated that from a mathematical
standpoint, a definition of the Dirac Hamiltonian as a s.a. operator presents
no problem for arbitrary $Z$. A specific feature of the overcritical charges
is a nonuniqueness of the s.a. Dirac Hamiltonian, but this nonuniqueness is
characteristic even for $Z>Z_{\mathrm{s}}=\left(  \sqrt{3}/2\right)
\alpha^{-1}\simeq118,68$. For each $Z\geq Z_{\mathrm{s}}$, there exist a
family of s.a. Dirac Hamiltonians parametrized by a finite number of extra
parameters\textrm{ }(and specified by additional boundary conditions at the
origin). The existence of these parameters is a manifestation of a nontrivial
physics inside the nucleus. A real spectrum and a complete set of eigenstates
can be evaluated for each Hamiltonian, so that a relativistic quantum
mechanics for an electron in such a Coulomb field can be constructed.

In the present article, we briefly summarize all the previously obtained
formal results in a form more suitable for physicists with some additional new
important details and numerical calculations of the electronic spectra. The
spectrum of each Hamiltonian consists of a universal\textrm{ }continuous part
that is a union of two intervals $(-\infty,m]$ and $[m,\infty)$ and a specific
discrete spectrum located in the interval $[-m,m)$. We concentrate on the
discrete spectrum\textrm{.}

\section{Dirac Hamiltonian with Coulomb field\label{S2}}

We consider an electron of charge $-e<0$ and mass $m$\ moving in the Coulomb
field of charge $Ze>0$. We describe this field by a scalar electromagnetic
potential of the form $A_{0}=Zer^{-1}$, we set $\hbar=c=1$ in what follows. A
behavior of the electron in the Coulomb field is governed by the Dirac
Hamiltonian $\hat{H}\left(  Z\right)  $ that is a s.a. operator in the Hilbert
space $\mathfrak{H}$ of square-integrable bispinors $\Psi\left(
\mathbf{r}\right)  .$ On its domain which must be properly specified, $\hat
{H}\left(  Z\right)  $ acts by the differential operation%
\begin{equation}
\check{H}\left(  Z\right)  =\gamma^{0}\left(  \boldsymbol{\gamma
}\mathbf{\check{p}}+m\right)  -qr^{-1},\ \mathbf{\check{p}}=-i\mathbf{\nabla
,\ }r=\left\vert \mathbf{r}\right\vert ,\ q=Z\alpha\label{1a}%
\end{equation}
(in what follows, we use $\gamma$-matrices in the standard representation). In
the problem under consideration, there are three commuting s.a. operators
$\boldsymbol{\hat{J}}^{2}$, $\hat{J}_{z}$, and $\hat{K}$, where
$\boldsymbol{\hat{J}}$ is the total\ angular\ momentum and $\hat{K}$ is the
so-called spin operator%
\[
\boldsymbol{\hat{J}}=\boldsymbol{\hat{L}}+\mathbf{\Sigma/}2,\ \boldsymbol{\hat
{L}}=\left[  \mathbf{r}\times\mathbf{\hat{p}}\right]  ,\ \hat{K}=\gamma
^{0}\left[  1+\left(  \boldsymbol{\Sigma\hat{L}}\right)  \right]  .
\]
All they commute with $\hat{H}\left(  Z\right)  $. Any bispinor $\Psi\left(
\mathbf{r}\right)  $ can be represented as $\Psi(\mathbf{r})=\sum_{j,M,\zeta
}\Psi_{j,M,\zeta}\left(  \mathbf{r}\right)  $, where $\Psi_{j,M,\zeta}$ are
bispinors of the form%
\begin{equation}
\Psi_{j,M,\zeta}\left(  \mathbf{r}\right)  =\frac{1}{r}\left(
\begin{array}
[c]{l}%
\Omega_{j,M,\zeta}(\theta,\varphi)f\left(  r\right) \\
i\Omega_{j,M,-\zeta}(\theta,\varphi)g\left(  r\right)
\end{array}
\right)  , \label{R.1a}%
\end{equation}
$\Omega_{j,M,\zeta}$ are normalized spherical spinors, $f\left(  r\right)  $
and $g\left(  r\right)  $ are radial functions, and $j=1/2,3/2,...$%
,\ $M=-j,-j+1,...,j$,\ $\zeta=\pm$. Bispinors $\Psi_{j,M,\zeta}$ are
eigenvectors of $\boldsymbol{\hat{J}}^{2}$, $\hat{J}_{z}$, and $\hat{K}$,
\begin{equation}
\boldsymbol{\hat{J}}^{2}\Psi=j(j+1)\Psi,\;\hat{J}_{z}\Psi=M\Psi,\ \hat{K}%
\Psi=-\zeta(j+1/2)\Psi\ .\nonumber
\end{equation}

Let $\mathbb{L}^{2}\left(  \mathbb{R}_{+}\right)  =L^{2}(\mathbb{R}_{+})\oplus
L^{2}(\mathbb{R}_{+})$ (where $L^{2}(\mathbb{R}_{+})$ is the space of
functions of $r$ square-integrable on the semiaxis $\mathbb{R}_{+}=[0,\infty
)$) be the Hilbert space of doublets $F(r)$,
\[
F(r)=\left(
\begin{array}
[c]{c}%
f(r)\\
g(r)
\end{array}
\right)  =\left(  f(r)\diagup g\left(  r\right)  \right)  ,
\]
with the scalar product
\[
\left(  F_{1},F_{2}\right)  =\int_{\mathbb{R}_{+}}drF_{1}^{+}\left(  r\right)
F_{2}\left(  r\right)  =\int_{\mathbb{R}_{+}}dr\left[  \overline{f_{1}\left(
r\right)  }f_{2}\left(  r\right)  +\overline{g_{1}\left(  r\right)  }%
g_{2}\left(  r\right)  \right]  .
\]
Then rep. (\ref{R.1a}) and the relation
\[
||\Psi_{j,M,\zeta}||^{2}=\int_{\mathbb{R}_{+}}dr\left[  |f(r)|^{2}%
+|g(r)|^{2}\right]
\]
show that any subspace of bispinors $\Psi_{j,M,\zeta}$ with fixed $j,M,\zeta$
is unitary equivalent to $\mathbb{L}^{2}\left(  \mathbb{R}_{+}\right)  $, an
explicit form of this equivalence is%
\begin{equation}
\Psi_{j,M,\zeta}(\mathbf{r})=r^{-1}\mathbf{\Pi}_{j,M,\zeta}(\theta
,\varphi)F(r),\ \ F(r)=r\int\sin\theta d\theta d\varphi\mathbf{\Pi}%
_{j,M,\zeta}^{+}(\theta,\varphi)\Psi_{j,M,\zeta}(\mathbf{r}). \label{Pi}%
\end{equation}
Here, $\mathbf{\Pi}_{j,M,\zeta}$ and $\mathbf{\Pi}_{j,M,\zeta}^{+}$ are the
respective $(4\times2)$- and $(2\times4)$-matrices,%
\begin{align*}
&  \mathbf{\Pi}_{j,M,\zeta}=\left(
\begin{array}
[c]{cc}%
\Omega_{j,M,\zeta} & \mathbf{0}\\
\mathbf{0} & i\Omega_{j,M,-\zeta}%
\end{array}
\right)  ,\ \mathbf{\Pi}_{j,M,\zeta}^{+}=\left(
\begin{array}
[c]{cc}%
\Omega_{j,M,\zeta}^{+} & \mathbf{0}^{T}\\
\mathbf{0}^{T} & -i\Omega_{j,M,-\zeta}^{+}%
\end{array}
\right)  ,\\
&  \int\sin\theta d\theta d\varphi\left[  \mathbf{\Pi}_{j,M,\zeta}^{+}%
(\theta,\varphi)\mathbf{\Pi}_{j,M,\zeta}(\theta,\varphi)\right]  _{ab}%
=\delta_{ab},\ \ a,b=1,2,
\end{align*}
where $\mathbf{0=}\left(  0\diagup0\right)  $ is a two-column and
$\mathbf{0}^{T}=\left(  0,0\right)  $ is a two-row. The stationary
Schr\"{o}dinger equation $\hat{H}\left(  Z\right)  \Psi\left(  \mathbf{r}%
\right)  =E\Psi\left(  \mathbf{r}\right)  $ in the Hilbert space
$\mathfrak{H}$\ is reduced to radial equations%
\[
\hat{h}\left(  Z,j,\zeta\right)  F\left(  r\right)  =EF\left(  r\right)
,\ \ F\in\mathbb{L}^{2}(\mathbb{R}_{+}),
\]
in the Hilbert space $\mathbb{L}^{2}(\mathbb{R}_{+})$, where $\hat{h}\left(
Z,j,\zeta\right)  $ are s.a. partial\textrm{ }radial Hamiltonians acting on
the doublets $F\left(  r\right)  $ by radial differential operations%
\begin{equation}
\check{h}\left(  Z,j,\zeta\right)  =-i\sigma_{2}d_{r}+\zeta(j+1/2)r^{-1}%
\sigma_{1}-qr^{-1}+m\sigma_{3}. \label{2}%
\end{equation}
The problem of constructing a rotationally invariant s.a. Dirac Hamiltonian
$\hat{H}\left(  Z\right)  $ is reduced to the problem of constructing s.a.
partial radial Hamiltonians $\hat{h}\left(  Z,j,\zeta\right)  $. We construct
all possible s.a. partial radial Hamiltonians $\hat{h}\left(  Z,j,\zeta
\right)  $ using \ the theory of s.a. extensions of symmetric operators, which
is reduced to specifying their domains\textrm{ }of definition $D_{\hat
{h}\left(  Z,j,\zeta\right)  }\subset$ $\mathbb{L}^{2}(\mathbb{R}_{+})$. By
construction, each operator $\hat{h}\left(  Z,j,\zeta\right)  $ is a s.a.
extension of the\textrm{ }so-called initial symmetric operator $\hat
{h}_{\mathrm{in}}\left(  Z,j,\zeta\right)  $ with the domain $D_{\hat
{h}_{\mathrm{in}}\left(  Z,j,\zeta\right)  }=\mathcal{D}(\mathbb{R}_{+}%
)\oplus\mathcal{D}(\mathbb{R}_{+})$, where $\mathcal{D}(\mathbb{R}_{+})$ is a
space of smooth compactly supported functions on the semiaxis $\mathbb{R}_{+}%
$, and\textrm{ }(simultaneously) is, generally, a\textrm{ }s.a. restriction of
the adjoint operator $\hat{h}_{\mathrm{in}}^{+}\left(  Z,j,\zeta\right)  $
defined on the so-called natural domains $D_{\check{h}\left(  Z,j,\zeta
\right)  }^{\ast}\left(  \mathbb{R}_{+}\right)  $ of doublets $F\in
\mathbb{L}^{2}(\mathbb{R}_{+})$ that are absolutely continuos in
$\mathbb{R}_{+}$ and such that $\check{h}\left(  Z,j,\zeta\right)
F\in\mathbb{L}^{2}(\mathbb{R}_{+})$. Thus,\textrm{ }$D_{\hat{h}_{\mathrm{in}%
}\left(  Z,j,\zeta\right)  }\subset\ D_{\hat{h}\left(  Z,j,\zeta\right)
}\subseteq D_{\check{h}\left(  Z,j,\zeta\right)  }^{\ast}\left(
\mathbb{R}_{+}\right)  $, and if $D_{\hat{h}\left(  Z,j,\zeta\right)  }$ does
not coincide with $D_{\check{h}\left(  Z,j,\zeta\right)  }^{\ast}\left(
\mathbb{R}_{+}\right)  $, it is specified by some additional asymptotic
boundary conditions at the origin, which are defined not uniquely.

A result of constructing s.a. radial Hamiltonians $\hat{h}\left(
Z,j,\zeta\right)  $ essentially depends on the values of the parameters $Z$
and $j$.

\begin{figure}[ptb]%
\centering
\includegraphics[
height=6.7907cm,
width=9.6295cm
]%
{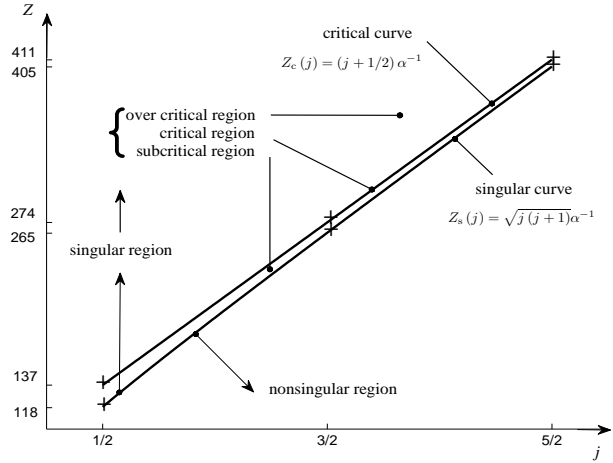}%
\caption{Regions in the $j,Z$ plane. Singular and critical curves.}%
\label{Fig. 1}%
\end{figure}

There are two regions in the first quadrant of the $j,Z$ plane, we call them
the nonsingular and singular ones, where the problem of s.a. extensions has
principally different solutions. These regions are separated by the singular
curve $Z=Z_{\mathrm{s}}\left(  j\right)  $, where%
\[
Z_{\mathrm{s}}\left(  j\right)  =\sqrt{j\left(  j+1\right)  }\alpha^{-1},
\]
so that the nonsingular and singular regions are defined by the respective
inequalities $Z\leq Z_{\mathrm{s}}\left(  j\right)  $ and $Z>Z_{\mathrm{s}%
}\left(  j\right)  $. The values%
\[
Z_{\mathrm{s}}\left(  1/2\right)  =118,68;\ \ Z_{\mathrm{s}}\left(
3/2\right)  =265,37;\ Z_{\mathrm{s}}\left(  5/2\right)  =405,36;\ ...\ ,
\]
can be called the singular values of $Z$ for a given $j,$ see FIG.
\ref{Fig. 1}.%

In what follows, we describe s.a. radial Hamiltonians $\hat{h}\left(
Z,j,\zeta\right)  $ and their spectra in the nonsingular and singular regions
separately. The consideration in the non-singular and singular regions is true
for the whole set of all quantum numbers.

All the mathematical details concerning s.a. extensions and a spectral
analysis of Hamiltonians\textrm{ }(including complete sets of
eigenfunctions)\textrm{ }based on the Krein method of guiding functionals can
be found in \cite{VorGiT07a,VorGiT,book}.

\section{Nonsingular region\label{S3}}

In the nonsingular region, $Z\leq Z_{\mathrm{s}}\left(  j\right)  $, the
partial radial Hamiltonian $\hat{h}\left(  Z,j,\zeta\right)  $ is defined
uniquely, $\hat{h}\left(  Z,j,\zeta\right)  =\hat{h}_{\mathrm{in}}^{+}\left(
Z,j,\zeta\right)  $, and its domain is\textrm{ }$D_{\hat{h}\left(
Z,j,\zeta\right)  }=D_{\check{h}\left(  Z,j,\zeta\right)  }^{\ast}\left(
\mathbb{R}_{+}\right)  $. The functions belonging to $D_{\check{h}\left(
Z,j,\zeta\right)  }^{\ast}\left(  \mathbb{R}_{+}\right)  $ have the following
asymptotic behavior
\[
F(r)=O(r^{1/2}),\;r\rightarrow0;\ F(r)\rightarrow0,\ r\rightarrow\infty.
\]
A discrete spectrum $\left\{  E_{n^{\left(  \zeta\right)  }}\left(
Z,j,\zeta\right)  \right\}  \,$of each Hamiltonian $\hat{h}\left(
Z,j,\zeta\right)  $ is given by%
\begin{equation}
E_{n^{\left(  \zeta\right)  }}\left(  Z,j,\zeta\right)  =\frac{m\left(
n^{\left(  \zeta\right)  }+\gamma\right)  }{\sqrt{q^{2}+(n^{\left(
\zeta\right)  }+\gamma)^{2}}},\ \gamma=\sqrt{\left(  j+1/2\right)  ^{2}-q^{2}%
}. \label{3a}%
\end{equation}
The quantum numbers $n^{\left(  \zeta\right)  }$ take the values%
\begin{equation}
n^{\left(  +\right)  }\in\mathbb{N}=\left\{  1,2,...\right\}  ;\ n^{\left(
-\right)  }\in\mathbb{Z}_{+}=\left\{  0,1,2,...\right\}  . \label{3b}%
\end{equation}
The expression (\ref{3a}) coincides with the well-known Sommerfeld formula for
the spectrum of the Dirac electron in the Coulomb field. This result justifies
the standard formal treatment of the Dirac Hamiltonian\ with $Z$ in the
nonsingular region in the physical literature where the Dirac Hamiltonian
is\ identified with the differential operation $\check{H}_{D}(Z)$ (\ref{1a})
and the natural domain is implicitly assumed.

\section{Singular region\label{S4}}

In the singular region, $Z>Z_{\mathrm{s}}\left(  j\right)  $, the s.a. radial
Hamiltonian $\hat{h}\left(  Z,j,\zeta\right)  ~$with given $Z,j,\zeta$ is
defined not uniquely (the deficiency indices of the operator $\hat
{h}_{\mathrm{in}}\left(  Z,j,\zeta\right)  $ are $\left(  1,1\right)  $).
There exists a one-parameter family$\{\hat{h}_{\nu}\left(  Z,j,\zeta\right)
,\ \nu\in\mathbb{[}-\pi/2,\pi/2],-\pi/2\sim\pi/2\}$ of Hamiltonians defined on
different domains $D_{\hat{h}_{\nu}\left(  Z,j,\zeta\right)  }\subset
D_{\check{h}\left(  q,\varkappa\right)  }^{\ast}\left(  \mathbb{R}_{+}\right)
$ specified by different asymptotic s.a. boundary conditions at the origin,
which are parametrized by the parameter $\nu$.

The position of the discrete energy levels $E_{n^{\left(  \zeta\right)  }%
}^{\left(  \nu\right)  }\left(  Z,j,\zeta\right)  $ essentially depends on
$\nu$, in particular, there exists a value $\nu=\nu_{-m}$, for which the lower
energy level coincides with the boundary $E=-m$ of the lower continuous spectrum.

Technically, it is convenient to divide the singular region into three
subregions, we call them subcritical, critical, and overcritical regions. The
subregions are distinguished by a character of asymptotic boundary conditions
at the origin.

The boundary conditions are similar in each subregion, which provides similar
solutions of the corresponding spectral problems. In what follows, we describe
these subregions, the domains $D_{\hat{h}_{\nu}\left(  Z,j,\zeta\right)  }$ in
these subregions, and some details of discrete spectra.

\subsection{Subcritical regions\label{S4.1}}

The subcritical region is defined by the inequalities $Z_{\mathrm{s}}\left(
j\right)  <Z<Z_{\mathrm{c}}\left(  j\right)  ,$ where%
\begin{equation}
Z_{\mathrm{c}}\left(  j\right)  =\left(  j+1/2\right)  \alpha^{-1}, \label{4}%
\end{equation}
see FIG. \ref{Fig. 1}. The values%
\[
Z_{\mathrm{c}}\left(  1/2\right)  =Z_{\mathrm{c}}=137,04;\ Z_{\mathrm{c}%
}\left(  3/2\right)  =274,08;\ Z_{\mathrm{c}}\left(  5/2\right)
=411,12;\ ...\
\]
can be called the critical $Z$-value for a given $j$. In the subcritical
region, the s.a. \ radial Hamiltonians $\hat{h}_{\nu}\left(  Z,j,\zeta\right)
$ are specified by s.a. boundary conditions,%
\begin{equation}
F(r)=c[(mr)^{\gamma}d_{+}\cos\nu+(mr)^{-\gamma}d_{-}\sin\nu]+O(r^{1/2}%
),\;r\rightarrow0, \label{4.1.1}%
\end{equation}
where $\gamma=\sqrt{\left(  j+1/2\right)  ^{2}-q^{2}}$,$\ 0<\gamma<1/2$, $c$
is an arbitrary complex number, and $d_{\pm}$ are some constant doublets.

The spectrum of\textrm{ }each $\hat{h}_{\nu}\left(  Z,j,\zeta\right)  $ is
simple (nondegenerate) and consists of a continuous part that is the set
$(-\infty,-m]\cup\lbrack m,\infty)$ and a discrete part located in the
interval $[-m,m)$. The discrete spectrum is a\ growing infinite sequence
$\{E_{n^{\left(  \zeta\right)  }}^{\left(  \nu\right)  }\left(  Z,j,\zeta
\right)  \}$ of the energy levels that are the roots of the equation%

\begin{align}
&  \frac{f(E)\cos\nu+\Gamma(1-2\gamma)\sin\nu}{f(E)\sin\nu-\Gamma
(1-2\gamma)\cos\nu}=0,\label{IIEn}\\
&  f\left(  E\right)  =\frac{\Gamma(1+2\gamma)\Gamma(-\gamma-qE\tau
^{-1})[q(m-E)-(\varkappa+\gamma)\tau]}{\Gamma(\gamma-qE/\tau
)[q(m-E)-(\varkappa-\gamma)\tau](2\tau/m)^{2\gamma}},\nonumber\\
&  \varkappa=\zeta(j+1/2),\;\tau=\sqrt{m^{2}-E^{2}},\nonumber
\end{align}
the integers $n^{\left(  \zeta\right)  }$ take the previous values
(\ref{3b})\textrm{.}

We outline the most important features of the discrete spectrum. For each
$Z,j,\zeta$, there exists $\nu=\nu_{-m}=\nu_{-m}(Z,j)$ (independent of $\zeta
$) such that the lowest energy level is equal to $-m$. This $\nu_{-m}$ is
determined from eq. (\ref{IIEn}) by setting $E=-m$ and noting that
$f(-m)=\Gamma(1+2\gamma)(2q)^{-2\gamma\text{ }}$ to yield
\[
\tan\nu_{-m}=-\frac{\Gamma(1+2\gamma)}{\Gamma(1-2\gamma)}(2q)^{-2\gamma\text{
}}%
\]
For fixed $n^{\left(  \zeta\right)  }$, $E_{n^{\left(  \zeta\right)  }%
}^{\left(  \nu\right)  }\left(  Z,j,\zeta\right)  $ as functions of $\nu$ are
monotonically decreasing functions with the properties%
\[
\lim_{\nu\rightarrow-\pi/2+0}E_{n^{\left(  \zeta\right)  }}^{\left(
\nu\right)  }\left(  Z,j,\zeta\right)  =\lim_{\nu\rightarrow\pi/2-0}%
E_{n^{\left(  \zeta\right)  }+1}^{\left(  \nu\right)  }\left(  Z,j,\zeta
\right)  \equiv\mathcal{E}_{n^{\left(  \zeta\right)  }}\ .
\]
Let%
\[
n_{0}^{\left(  \zeta\right)  }=n_{\min}^{(\zeta)}=\left\{
\begin{array}
[c]{l}%
1,\ \zeta=+\\
0,\ \zeta=-
\end{array}
\right.  .
\]
A subtlety is that the function $E_{n_{0}^{\left(  \zeta\right)  }}^{\left(
\nu\right)  }\left(  Z,j,\zeta\right)  $ is defined only for $\nu\in
\lbrack-\pi/2,\nu_{-m}]$, which implies that in the energy interval $\left[
-m,\mathcal{E}_{n_{0}^{\left(  \zeta\right)  }}\right)  $, there are no energy
level for $\nu\in(\nu_{-m},\pi/2]$, while for each $\nu\in\lbrack-\pi
/2,\nu_{-m}]$, there is one level $E_{n_{0}^{\left(  \zeta\right)  }}^{\left(
\nu\right)  }\left(  Z,j,\zeta\right)  $ monotonically growing from $-m$ to
$\mathcal{E}_{n_{0}^{\left(  \zeta\right)  }}$ when $\nu$ changes from
$\nu_{-m}$ to $-\pi/2$.

The functions $E_{n^{\left(  \zeta\right)  }}^{\left(  \nu\right)  }\left(
Z,j,\zeta\right)  ~$with $n^{\left(  \zeta\right)  }>n_{0}^{\left(
\zeta\right)  }$ are defined for all $\nu\in\lbrack-\pi/2,\pi/2]$. In each
energy interval $\left[  \mathcal{E}_{n^{\left(  \zeta\right)  }}%
,\mathcal{E}_{n^{\left(  \zeta\right)  }+1}\right]  $, there is one level
$E_{n^{\left(  \zeta\right)  }+1}^{\left(  \nu\right)  }\left(  Z,j,\zeta
\right)  $ monotonically growing from $\mathcal{E}_{n^{\left(  \zeta\right)
}}$ to $\mathcal{E}_{n^{\left(  \zeta\right)  }+1}$ when $\nu$ changes from
$\pi/2$ to $-\pi/2$. We note that the states with the energies $E_{n^{\left(
\zeta\right)  }}^{\left(  -\pi/2\right)  }\left(  Z,j,\zeta\right)  $ and
$E_{n^{\left(  \zeta\right)  }+1}^{\left(  \pi/2\right)  }\left(
Z,j,\zeta\right)  =E_{n^{\left(  \zeta\right)  }}^{\left(  -\pi/2\right)
}\left(  Z,j,\zeta\right)  =\mathcal{E}_{n^{\left(  \zeta\right)  }}$
represent the same eigenstate. It follows from the fact that according to eq.
(\ref{4.1.1}), the values $\nu=\pi/2$ and $\nu=-\pi/2$ are equivalent and
therefore the Hamiltonians $\hat{h}_{\pi/2}\left(  Z,j,\zeta\right)  $ and
$\hat{h}_{-\pi/2}\left(  Z,j,\zeta\right)  $ are the same. The eigenvalues
$\mathcal{E}_{n^{\left(  \zeta\right)  }}$ can be found explicitly. For
$\nu=\pm\pi/2$, eq. (\ref{IIEn}) is reduced to the equation $1/f\left(
E\right)  =0$, and we find%
\[
\mathcal{E}_{n^{\left(  \zeta\right)  }}=\frac{(n^{\left(  \zeta\right)
}-\gamma)m}{\sqrt{q^{2}+(n^{\left(  \zeta\right)  }-\gamma)^{2}}}.
\]

In particular, we see that the discrete spectrum $E_{n^{\left(  \zeta\right)
}}^{\left(  0\right)  }\left(  Z,j,\zeta\right)  $ is given by eq. (\ref{3a})
with $Z$ corresponding to the region under consideration.

\begin{figure}[ptbh]%
\centering
\includegraphics[
height=5.8079cm,
width=7.9298cm
]%
{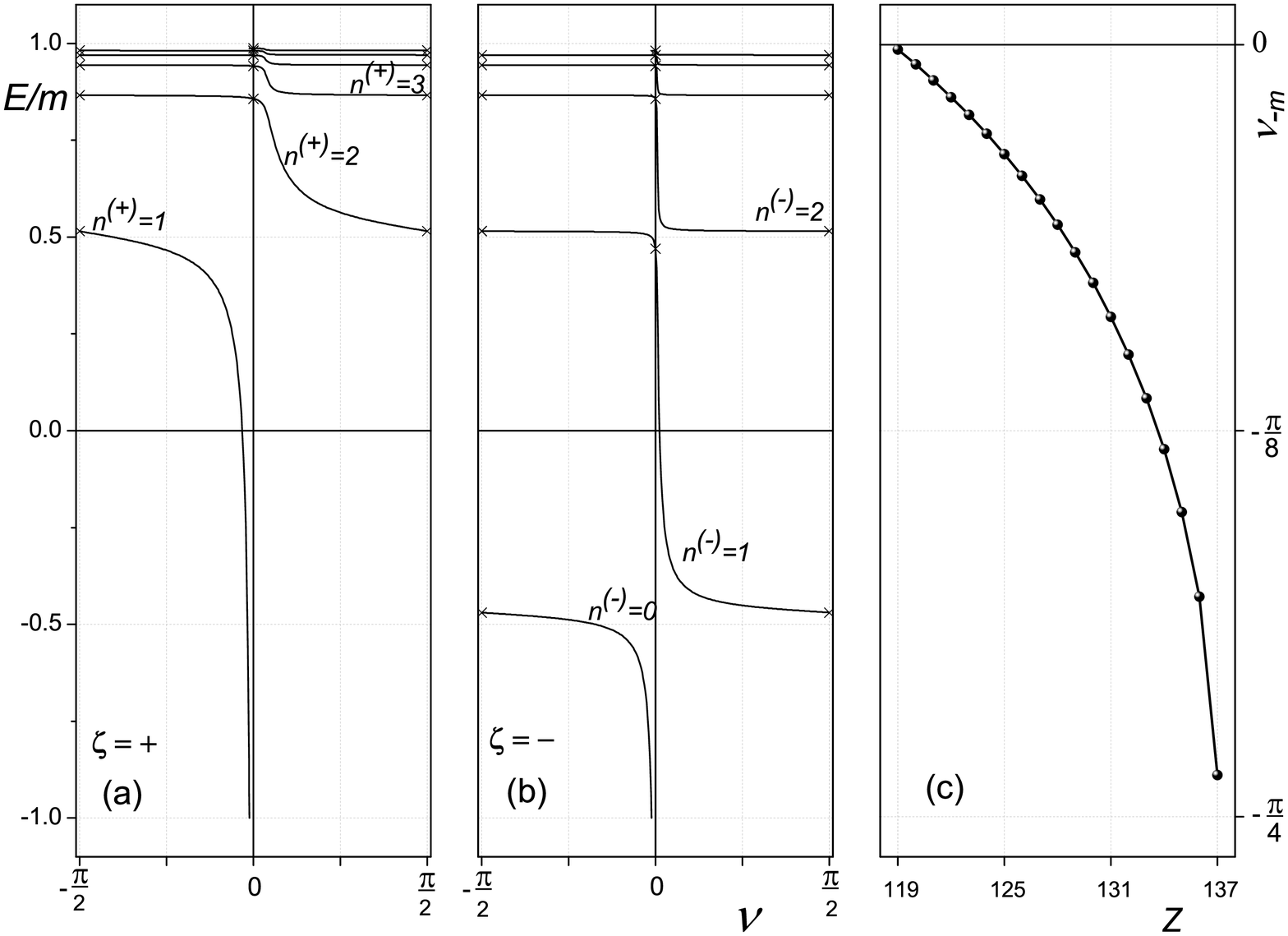}%
\caption{$\nu$-dependence of energy levels $E_{n^{\left(  \zeta\right)  }%
}^{\left(  \nu\right)  }\left(  121,1/2,\zeta=\pm\right)  $ and $Z$-dependence
of $\left.  \nu_{-m}\right\vert _{j=1/2}.$}%
\label{Fig. 2}%
\end{figure}

For illustration, we give graphs of five low energy levels ($Z=121$, $j=1/2$)
as functions of $\nu$, for $\zeta=+$ (FIG. \ref{Fig. 2}a), for $\zeta=-$ (FIG.
\ref{Fig. 2}b), and also a graph of the parameter $\left.  \nu_{-m}\right\vert
_{j=1/2}$ as a function of $Z,$ see FIG. \ref{Fig. 2}c.

In addition, to give an idea of energy scale, we give Table 1 of numerical
values of some energy levels.

\begin{center}
{Table 1. Energy levels for $Z=121,\ j=1/2$\newline}
{\scriptsize \addtolength{\tabcolsep}{-1pt}
\begin{tabular}
[c]{|c||c|c||c|c||c|c||c|c||c|c|}\hline\hline
$\nu$ & \multicolumn{2}{|c||}{$\mathbf{E}_{0}$} &
\multicolumn{2}{|c||}{$\mathbf{E}_{1}$} & \multicolumn{2}{|c||}{$\mathbf{E}%
_{2}$} & \multicolumn{2}{|c||}{$\mathbf{E}_{3}$} &
\multicolumn{2}{|c|}{$\mathbf{E}_{4}$}\\\hline\hline
-$\frac{\pi}{2}$ &  & $-0.469411$ & $0.515067$ & $0.515067$ & $0.866198$ &
$0.866198$ & $0.944175$ & $0.944175$ & $0.970121$ & $0.970121$\\\hline
-$\frac{\pi}{4}$ &  & $-0.487899$ & $0.465810$ & $0.514264$ & $0.865181$ &
$0.866078$ & $0.943927$ & $0.944142$ & $0.970026$ & $0.970108$\\\hline
$0$ &  &  &  & $0.46941$ & $0.85715$ & $0.85715$ & $0.941615$ & $0.941615$ &
$0.969107$ & $0.941615$\\\hline
$\frac{\pi}{4}$ &  &  &  & $-0.450991$ & $0.563241$ & $0.515898$ & $0.867497$
& $0.866322$ & $0.944482$ & $0.944209$\\\hline
$\frac{\pi}{2}$ &  &  &  & $-0.469411$ & $0.515067$ & $0.515067$ & $0.866198$
& $0.866198$ & $0.944175$ & $0.944175$\\\hline\hline
$\zeta$ & $+$ & $-$ & $+$ & $-$ & $+$ & $-$ & $+$ & $-$ & $+$ & $-$%
\\\hline\hline
\end{tabular}
}
\end{center}

\subsection{Critical region\label{S4.2}}

The critical region is the critical curve $Z=Z_{\mathrm{c}}\left(  j\right)
,$ see FIG. \ref{Fig. 1}. For integer $Z$, this region does not exist if the
fine structure constant $\alpha$ is an irrational number, see (\ref{4}). In
particular, this region certainly is absent for $j=1/2$. In the critical
region, the s.a. radial Hamiltonian $\hat{h}_{\nu}\left(  Z,j,\zeta\right)  $
is specified by s.a. boundary conditions at the origin of the form%
\begin{equation}
F(r)=c\left[  d_{0}(r)\cos\nu+d_{+}\sin\nu\right]  +O(r^{1/2}\ln
r),\;r\rightarrow0, \label{4.2.1}%
\end{equation}
where $d_{0}(r)$ are some doublet with the asymptotic behavior $d_{0}%
(r)=O\left(  \ln mr\right)  $ as$\ r\rightarrow0$.

The spectrum of\textrm{ }each $\hat{h}_{\nu}\left(  Z,j,\zeta\right)  $ is
simple (nondegenerate) and consists of a continuous part that is the set
$(-\infty,-m]\cup\lbrack m,\infty)$ and a discrete part located in the
interval $[-m,m)$. The discrete spectrum is a\ growing infinite sequence
$\{E_{n^{\left(  \zeta\right)  }}^{\left(  \nu\right)  }\left(  Z,j,\zeta
\right)  \}$ of energy levels that are the roots of the equation%
\begin{align}
&  \frac{g(E)\cos\nu-\sin\nu}{g(E)\sin\nu+\cos\nu}=0,\ \ g(E)=\ln
(2\tau/m)\nonumber\\
&  \ +\psi(-\left(  j+1/2\right)  E/\tau)+\frac{\zeta-\left(  \tau+\zeta
m\right)  /E}{2\left(  j+1/2\right)  }-2\psi(1), \label{IIIEn}%
\end{align}
where $\psi(x)=\Gamma^{\prime}(x)/\Gamma(x)$.

We outline the most important features of the discrete spectrum. For each
$Z,j,\zeta$, there exists $\nu=\nu_{-m}=\nu_{-m}(Z,j)$ (independent of $\zeta
$) such that the lowest energy level is equal to $-m$. This $\nu_{-m}$ is
determined from eq. (\ref{IIIEn}) by setting $E=-m$ and noting that
$g(-m)=\ln(2q_{\mathrm{c}})-2\psi(1)+\zeta/q_{\mathrm{c}}$ to yield
\[
\tan\nu_{-m}=\ln(2q_{\mathrm{c}})-2\psi(1)+\zeta/q_{\mathrm{c}}.
\]
For fixed $n^{\left(  \zeta\right)  }$, $E_{n^{\left(  \zeta\right)  }%
}^{\left(  \nu\right)  }\left(  Z,j,\zeta\right)  $ as functions of $\nu$ are
monotonically decreasing functions with the properties%
\[
\lim_{\nu\rightarrow-\pi/2+0}E_{n^{\left(  \zeta\right)  }}^{\left(
\nu\right)  }\left(  Z,j,\zeta\right)  =\lim_{\nu\rightarrow\pi/2-0}%
E_{n^{\left(  \zeta\right)  }+1}^{\left(  \nu\right)  }\left(  Z,j,\zeta
\right)  \equiv\mathcal{E}_{n^{\left(  \zeta\right)  }}.
\]

\begin{figure}[ptbh]%
\centering
\includegraphics[
height=5.5438cm,
width=7.9287cm
]%
{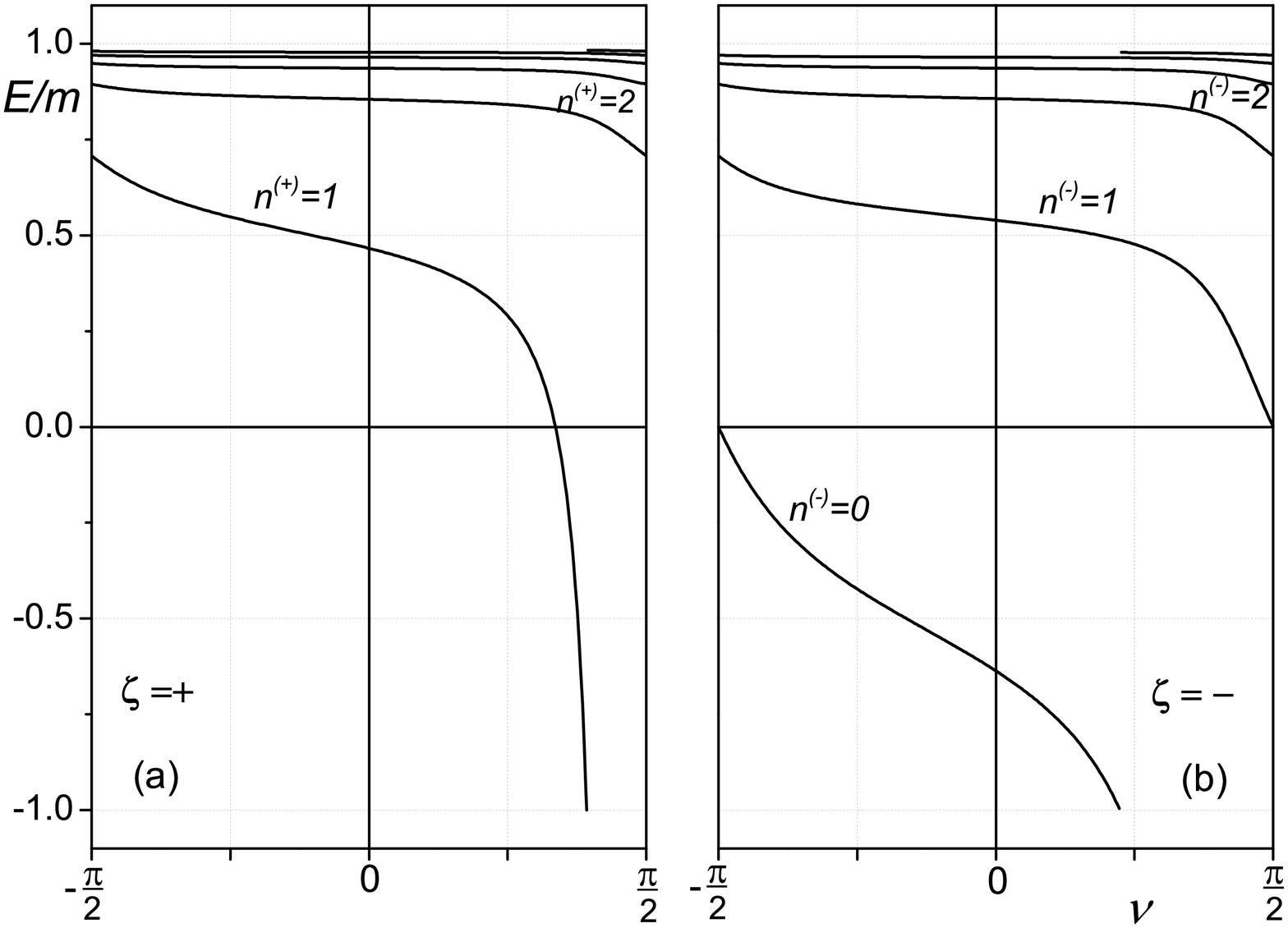}%
\caption{$\nu$-dependence of energy levels $E_{n^{\left(  \zeta\right)  }%
}^{\left(  \nu\right)  }\left(  Z_{c},1/2,\zeta=\pm\right)  $.}%
\label{Fig. c}%
\end{figure}

A subtlety is that the function $E_{n_{0}^{\left(  \zeta\right)  }}^{\left(
\nu\right)  }\left(  Z,j,\zeta\right)  $ is defined only for $\nu\in
\lbrack-\pi/2,\nu_{-m}]$, which implies that in the energy interval $\left[
-m,\mathcal{E}_{n_{0}^{\left(  \zeta\right)  }}\right)  $, there are no energy
level for $\nu\in(\nu_{-m},\pi/2]$, while for each $\nu\in\lbrack-\pi
/2,\nu_{-m}]$, there is one level $E_{n_{0}^{\left(  \zeta\right)  }}^{\left(
\nu\right)  }\left(  Z,j,\zeta\right)  $ monotonically growing from $-m$ to
$\mathcal{E}_{n_{0}^{\left(  \zeta\right)  }}$ when $\nu$ changes from
$\nu_{-m}$ to $-\pi/2$. The functions $E_{n^{\left(  \zeta\right)  }}^{\left(
\nu\right)  }\left(  Z,j,\zeta\right)  ~$with $n^{\left(  \zeta\right)
}>n_{0}^{\left(  \zeta\right)  }$ are defined for all $\nu\in\lbrack-\pi
/2,\pi/2]$. In each energy interval $\left[  \mathcal{E}_{n^{\left(
\zeta\right)  }},\mathcal{E}_{n^{\left(  \zeta\right)  }+1}\right]  $, there
is one level $E_{n^{\left(  \zeta\right)  }+1}^{\left(  \nu\right)  }\left(
Z,j,\zeta\right)  $ monotonically growing from $\mathcal{E}_{n^{\left(
\zeta\right)  }}$ to $\mathcal{E}_{n^{\left(  \zeta\right)  }+1}$ when $\nu$
changes from $\pi/2$ to $-\pi/2$. We note that the states with the energies
$E_{n^{\left(  \zeta\right)  }}^{\left(  -\pi/2\right)  }\left(
Z,j,\zeta\right)  $ and $E_{n^{\left(  \zeta\right)  }+1}^{\left(
\pi/2\right)  }\left(  Z,j,\zeta\right)  =E_{n^{\left(  \zeta\right)  }%
}^{\left(  -\pi/2\right)  }\left(  Z,j,\zeta\right)  =\mathcal{E}_{n^{\left(
\zeta\right)  }}$ represent the same eigenstate. It follows from the fact that
according to eq. (\ref{4.2.1}), the values $\nu=\pi/2$ and $\nu=-\pi/2$ are
equivalent and therefore the Hamiltonians $\hat{h}_{\pi/2}\left(
Z,j,\zeta\right)  $ and $\hat{h}_{-\pi/2}\left(  Z,j,\zeta\right)  $ are the
same. The eigenvalues $\mathcal{E}_{n^{\left(  \zeta\right)  }}$ can be found
explicitly. For $\nu=\pm\pi/2$, eq. (\ref{IIIEn}) is reduced to the equation
$1/g\left(  E\right)  =0$, and we find%
\begin{equation}
\mathcal{E}_{n^{\left(  \zeta\right)  }}=\frac{mn^{\left(  \zeta\right)  }%
}{\sqrt{\left(  j+1/2\right)  ^{2}+\left(  n^{\left(  \zeta\right)  }\right)
^{2}}}. \label{19}%
\end{equation}
For illustration, we give graphs of five low energy levels ($j=1/2$) as
functions of $\nu$, for $\zeta=+$ (FIG. \ref{Fig. c}a), for $\zeta=-$ (FIG.
\ref{Fig. c}b). In addition, to give an idea of energy scale, we give Table 2
of numerical values of some energy levels.%

\begin{center}
Table 2. Energy levels for $Z_{c}${$,\ j=1/2$}\newline%
{\normalsize {\scriptsize \addtolength{\tabcolsep}{-1pt}
\begin{tabular}
[c]{|c||c|c||c|c||c|c||c|c||c|c|}\hline\hline
$\nu$ & \multicolumn{2}{|c||}{$\mathbf{E}_{0}$} &
\multicolumn{2}{|c||}{$\mathbf{E}_{1}$} & \multicolumn{2}{|c||}{$\mathbf{E}%
_{2}$} & \multicolumn{2}{|c||}{$\mathbf{E}_{3}$} &
\multicolumn{2}{|c|}{$\mathbf{E}_{4}$}\\\hline\hline
-$\frac{\pi}{2}$ &  & $0$ & $0.707107$ & $0.707107$ & $0.894427$ & $0.894427$
& $0.948683$ & $0.948683$ & $0.970143$ & $0.970143$\\\hline
-$\frac{\pi}{4}$ &  & $0.476777$ & $0.548043$ & $0.581793$ & $0.864486$ &
$0.939071$ & $0.93892$ & $0.965944$ & $0.96591$ & $0.978405$\\\hline
$0$ &  & $-0.636533$ & $0.466006$ & $0.539248$ & $0.854923$ & $0.856797$ &
$0.936069$ & $0.936298$ & $0.964727$ & $0.964778$\\\hline
$\frac{\pi}{4}$ &  &  & $0.290246$ & $0.476777$ & $0.843971$ & $0.803545$ &
$0.932388$ & $0.920568$ & $0.96309$ & $0.963161$\\\hline
$\frac{\pi}{2}$ &  &  &  & $0$ & $0.707107$ & $0.707107$ & $0.894427$ &
$0.894427$ & $0.948683$ & $0.948683$\\\hline\hline
$\zeta$ & $+$ & $-$ & $+$ & $-$ & $+$ & $-$ & $+$ & $-$ & $+$ & $-$%
\\\hline\hline
\end{tabular}
}}
\end{center}

\subsection{Overcritical region\label{S4.3}}

The overcritical region is defined by the inequality $Z>Z_{\mathrm{c}}\left(
j\right)  $, $Z_{\mathrm{c}}\left(  j\right)  =\left(  j+1/2\right)
\alpha^{-1}$.

In this region, the s.a. radial Hamiltonians $\hat{h}_{\nu}\left(
Z,j,\zeta\right)  $ are specified by s.a. boundary conditions
\begin{equation}
F(r)=c\left[  i\mathrm{e}^{i\nu}(mr)^{i\sigma}\rho_{+}-i\mathrm{e}^{-i\nu
}(mr)^{-i\sigma}\rho_{-}\right]  +O(r^{1/2}),\;r\rightarrow0, \label{4.3.1}%
\end{equation}
where $\sigma=\sqrt{q^{2}-\left(  j+1/2\right)  ^{2}}$, $c$ is an arbitrary
complex number, and $\rho_{\pm}$ are some constant doublets.

The spectrum of\textrm{ }each $\hat{h}_{\nu}\left(  Z,j,\zeta\right)  $ is
simple (nondegenerate) and consists of a continuous part that is the set
$(-\infty,-m]\cup\lbrack m,\infty)$ and a discrete part located in the
interval $[-m,m)$. The discrete spectrum is is a\ growing infinite sequence
$\{E_{n}^{\left(  \nu\right)  }\left(  Z,j,\zeta\right)  \}$, $n\in
\mathbb{Z}_{+}$, of energy levels that are the roots of the equation%
\begin{align}
&  \cos\left[  \Theta(E)-\nu\right]  =0,\label{IVEn}\\
&  \Theta(E)=\frac{1}{2i}\sum_{a=1}^{3}\left[  \ln B_{a}-\left(  \ln
B_{a}\right)  ^{\ast}\right]  +\sigma\ln\frac{2\tau}{m},\nonumber
\end{align}
where $B_{1}=-2i\sigma$,$\ B_{2}\left(  E\right)  =i\sigma-Eq\tau^{-1}$,\ and
$B_{3}\left(  E\right)  =\tau(j+1/2-i\zeta\sigma)-\zeta q(m-E)$.

We outline the most important features of the discrete spectrum. For each
$Z,j,\zeta$, there exists $\nu=\nu_{-m}=\nu_{-m}(Z,j)$ (independent of $\zeta
$) such that the lowest energy level is equal to $-m$. This $\nu_{-m}$ is
determined from eq. (\ref{IVEn}) by setting $E=-m$ which yields%
\[
e^{2i(\nu_{-m}-\pi/2)}=e^{2i\Theta(-m)}=\frac{\Gamma(-2i\sigma)}%
{\Gamma(2i\sigma)}(2q)^{2i\sigma\text{ }}.
\]

The energy levels are determined by
\[
\Theta\left(  -m\right)  -\Theta\left(  E_{n}^{\left(  \nu\right)  }\left(
Z,j,\zeta\right)  \right)  =\pi n+\nu_{-m}-\nu.
\]
For fixed $n$, $E_{n}^{\left(  \nu\right)  }\left(  Z,j,\zeta\right)  $ as
functions of $\nu$ are monotonically decreasing functions with the properties%
\[
\lim_{\nu\rightarrow-\pi/2+0}E_{n}^{\left(  \nu\right)  }\left(
Z,j,\zeta\right)  =\lim_{\nu\rightarrow\pi/2-0}E_{n+1}^{\left(  \nu\right)
}\left(  Z,j,\zeta\right)  .
\]

A subtlety is that the function $E_{0}^{\left(  \nu\right)  }\left(
Z,j,\zeta\right)  $ is defined only for $\nu\in\lbrack-\pi/2,\nu_{-m}]$, which
implies that in the energy interval $\left[  -m,E_{0}^{\left(  -\pi/2\right)
}\left(  Z,j,\zeta\right)  \right)  $, there are no energy level for $\nu
\in(\nu_{-m},\pi/2]$, while for each $\nu\in\lbrack-\pi/2,\nu_{-m}]$, there is
one level $E_{0}^{\left(  \nu\right)  }\left(  Z,j,\zeta\right)  $
monotonically growing from $-m$ to $E_{0}^{\left(  -\pi/2\right)  }\left(
Z,j,\zeta\right)  $ when $\nu$ changes from $\nu_{-m}$ to $-\pi/2$.

The functions $E_{n}^{\left(  \nu\right)  }\left(  Z,j,\zeta\right)  ~$with
$n\geq1$ are defined for all $\nu\in\lbrack-\pi/2,\pi/2]$. In each energy
interval $\left[  E_{n}^{\left(  \pi/2\right)  }\left(  Z,j,\zeta\right)
=E_{n-1}^{\left(  -\pi/2\right)  }\left(  Z,j,\zeta\right)  ,E_{n}^{\left(
-\pi/2\right)  }\left(  Z,j,\zeta\right)  \right]  $, there is one level
$E_{n}^{\left(  \nu\right)  }\left(  Z,j,\zeta\right)  $ monotonically growing
from $E_{n}^{\left(  \pi/2\right)  }\left(  Z,j,\zeta\right)  =E_{n-1}%
^{\left(  -\pi/2\right)  }\left(  Z,j,\zeta\right)  $ to $E_{n}^{\left(
-\pi/2\right)  }\left(  Z,j,\zeta\right)  $ when $\nu$ changes from $\pi/2$ to
$-\pi/2$. We note that the states with the energies $E_{n}^{\left(
-\pi/2\right)  }\left(  Z,j,\zeta\right)  $ and $E_{n+1}^{\left(
\pi/2\right)  }\left(  Z,j,\zeta\right)  $ represent the same eigenstate. It
follows from the fact that according to eq. (\ref{4.3.1}), the values $\nu
=\pi/2$ and $\nu=-\pi/2$ are equivalent and therefore the Hamiltonians
$\hat{h}_{\pi/2}\left(  Z,j,\zeta\right)  $ and $\hat{h}_{-\pi/2}\left(
Z,j,\zeta\right)  $ are the same.

We note that in contrast to previous regions, the integers $n$ take the same
values for both $\zeta=+$ and $\zeta=-$ because there is no the Sommerfeld degeneracy.

\begin{figure}[ptbh]%
\centering
\includegraphics[
height=5.8079cm,
width=7.9276cm
]%
{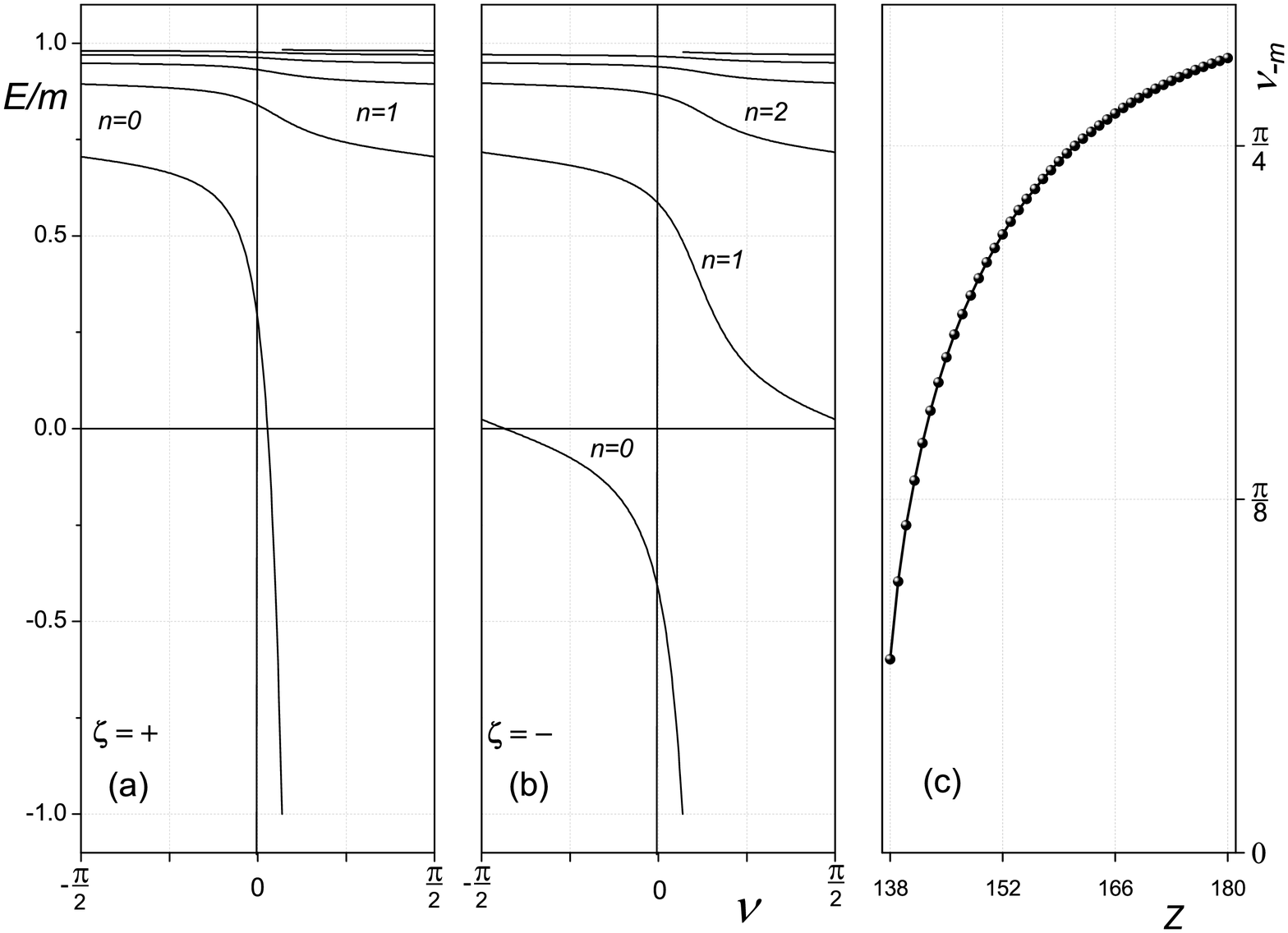}%
\caption{$\nu$-dependence of energy levels $E_{n}^{\left(  \nu\right)
}\left(  138,1/2,\zeta=\pm\right)  $ and $Z$-dependence of $\left.  \nu
_{-m}\right\vert _{j=1/2}.$}%
\label{Fig. 4}%
\end{figure}

For illustration, we give graphs of five low energy levels ($Z=138$, $j=1/2$)
as functions of $\nu$, for $\zeta=+$ (FIG. \ref{Fig. 4}a), for $\zeta=-$ (FIG.
\ref{Fig. 4}b), and also a graph of the parameter $\left.  \nu_{-m}\right\vert
_{j=1/2}$ as a function of $Z,$ see FIG. \ref{Fig. 4}c. In addition, to give
an idea of energy scale, we give Table 3 of numerical values of some energy
levels%

\begin{center}
Table 3. Energy levels for $Z=138${$,\ j=1/2$}\newline%
{\normalsize {\scriptsize \addtolength{\tabcolsep}{-1pt}
\begin{tabular}
[c]{|c||c|c||c|c||c|c||c|c||c|c|}\hline\hline
$\nu$ & \multicolumn{2}{|c||}{$\mathbf{E}_{0}$} &
\multicolumn{2}{|c||}{$\mathbf{E}_{1}$} & \multicolumn{2}{|c||}{$\mathbf{E}%
_{2}$} & \multicolumn{2}{|c||}{$\mathbf{E}_{3}$} &
\multicolumn{2}{|c|}{$\mathbf{E}_{4}$}\\\hline\hline
-$\frac{\pi}{2}$ & $0.705525$ & $0.024086$ & $0.893927$ & $0.717081$ &
$0.948327$ & $0.89653$ & $0.969889$ & $0.949229$ & $0.980395$ & $0.970294$%
\\\hline
-$\frac{\pi}{4}$ & $0.663242$ & $-0.075165$ & $0.884039$ & $0.682943$ &
$0.944912$ & $0.887947$ & $0.968366$ & $0.946203$ & $0.979595$ &
$0.96893$\\\hline
$0$ & $0.281658$ & $-0.42001$ & $0.839196$ & $0.584008$ & $0.931198$ &
$0.865425$ & $0.931198$ & $0.938723$ & $0.976677$ & $0.965675$\\\hline
$\frac{\pi}{4}$ &  &  & $0.742474$ & $0.165125$ & $0.903704$ & $0.764448$ &
$0.951885$ & $0.909191$ & $0.971521$ & $0.95389$\\\hline
$\frac{\pi}{2}$ &  &  & $0.705525$ & $0.024086$ & $0.893927$ & $0.717081$ &
$0.948327$ & $0.89653$ & $0.969889$ & $0.949229$\\\hline\hline
$\zeta$ & $+$ & $-$ & $+$ & $-$ & $+$ & $-$ & $+$ & $-$ & $+$ & $-$%
\\\hline\hline
\end{tabular}
}}

Table 4. Energy levels for $Z=180${$,\ j=1/2$}
{\normalsize {\scriptsize \addtolength{\tabcolsep}{-1pt}
\begin{tabular}
[c]{|c||c|c||c|c||c|c||c|c||c|c|}\hline\hline
$\nu$ & \multicolumn{2}{|c||}{$\mathbf{E}_{0}$} &
\multicolumn{2}{|c||}{$\mathbf{E}_{1}$} & \multicolumn{2}{|c||}{$\mathbf{E}%
_{2}$} & \multicolumn{2}{|c||}{$\mathbf{E}_{3}$} &
\multicolumn{2}{|c|}{$\mathbf{E}_{4}$}\\\hline\hline
-$\frac{\pi}{2}$ & $0.552275$ & $0.233202$ & $0.835423$ & $0.742113$ &
$0.917366$ & $0.883988$ & $0.95098$ & $0.936032$ & $0.967737$ & $0.959912$%
\\\hline
-$\frac{\pi}{4}$ & $0.363492$ & $-0.033728$ & $0.796897$ & $0.670614$ &
$0.903801$ & $0.861427$ & $0.944832$ & $0.92676$ & $0.964478$ & $0.955329$%
\\\hline
$0$ & $-0.005928$ & $-0.384837$ & $0.743034$ & $0.572097$ & $0.886487$ &
$0.832513$ & $0.937382$ & $0.91555$ & $0.960658$ & $0.967177$\\\hline
$\frac{\pi}{4}$ & $-0.83722$ & $-0.911674$ & $0.666155$ & $0.432269$ &
$0.864211$ & $0.794284$ & $0.928328$ & $0.901662$ & $0.956183$ &
$0.943688$\\\hline
$\frac{\pi}{2}$ &  &  & $0.552275$ & $0.233202$ & $0.835423$ & $0.742113$ &
$0.917366$ & $0.883988$ & $0.95098$ & $0.936032$\\\hline\hline
$\zeta$ & $+$ & $-$ & $+$ & $-$ & $+$ & $-$ & $+$ & $-$ & $+$ & $-$%
\\\hline\hline
\end{tabular}
}}
\end{center}

\begin{figure}[ptbh]%
\centering
\includegraphics[
height=5.8079cm,
width=7.9298cm
]%
{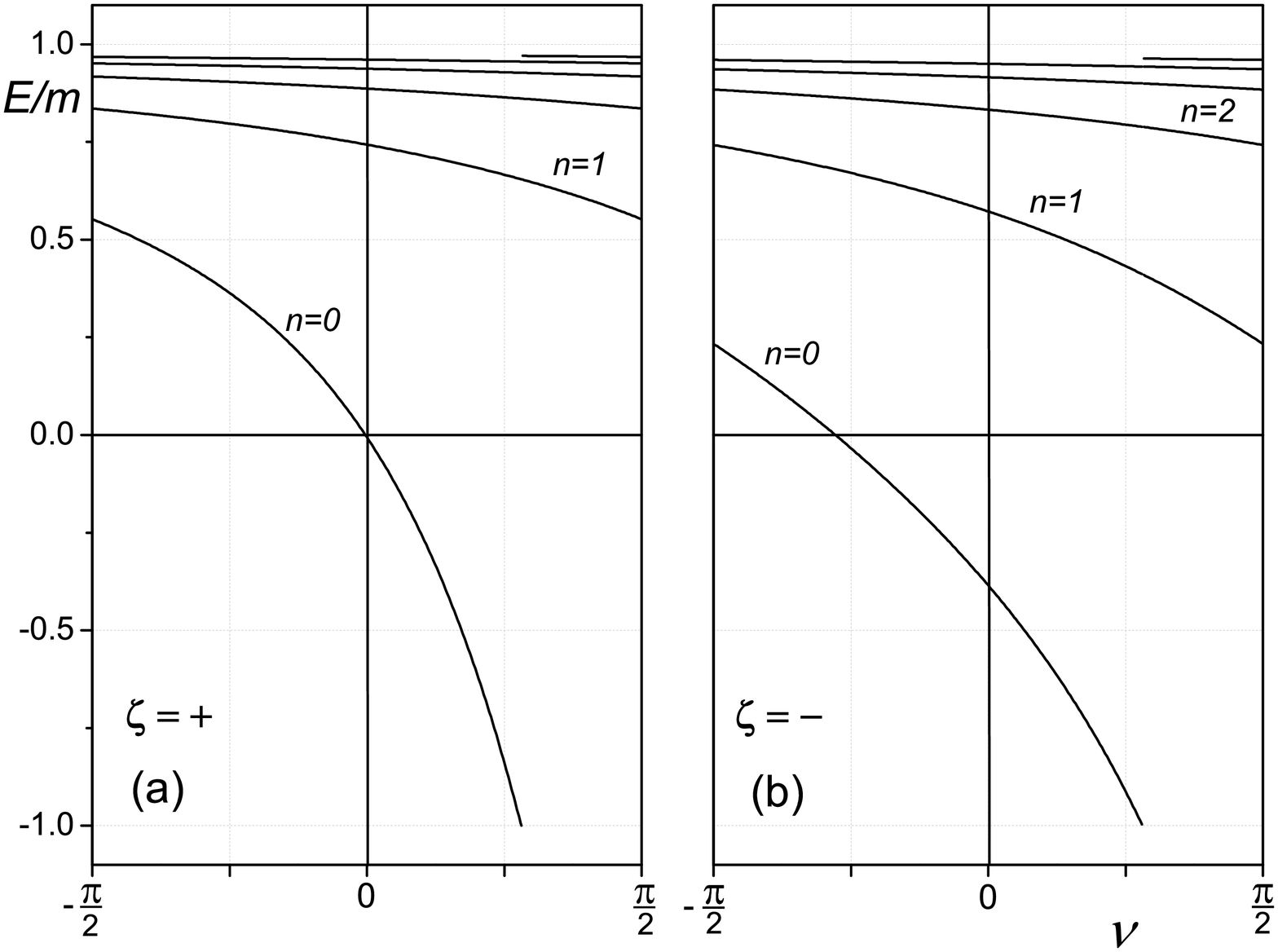}%
\caption{$\nu$-dependence of energy levels $E_{n}^{\left(  \nu\right)
}\left(  180,1/2,\zeta=\pm\right)  $.}%
\label{Fig. 5}%
\end{figure}

For comparison, we give graphs of five low energy levels ($Z=180$, $j=1/2$) as
functions of $\nu$, for $\zeta=+$ (FIG. \ref{Fig. 5}a), for $\zeta=-$ (FIG.
\ref{Fig. 5}b). In addition, to give an idea of energy scale, we give a Table
4 of numerical values of some energy levels.%

\section{Concluding remarks}

Here, we list the properties of all the radial s.a. Hamiltonians which are
common to both nonsingular and singular regions.

i) The spectrum of each s.a. Hamiltonian $\hat{h}\left(  Z,j,\zeta\right)  $
is simple (nondegenerate) and contains a continuous part that is the set
$(-\infty,-m]\cup\lbrack m,\infty)$ and a discrete part $\{E_{n}\left(
Z,j,\zeta\right)  \}$ located in the interval $[-m,m)$; for a precise meaning
of a nonnegative integer $n$, see the corresponding subsection.

ii) The discrete spectrum is always accumulated at the point $E=m$, and the
asymptotic form of the difference $E_{n}^{\mathrm{nonrel}}=$ $E_{n}\left(
Z,j,\zeta\right)  -m$ as $n\rightarrow\infty$ is given by the well-known
nonrelativistic formula%
\[
E_{n}^{\mathrm{nonrel}}=-mq^{2}\left(  2n^{2}\right)  ^{-1},
\]
the nonrelativistic spectrum does not depend on the extension parameter, $j$
and $\zeta$ (is degenerate in $j$ and $\zeta$).

iii) Eigenfunctions of the discrete spectrum and generalized eigenfunctions of
the continuous spectrum form a complete orthonormalized system in
$\mathbb{L}^{2}(\mathbb{R}_{+})$.

As soon as all s.a. radial Hamiltonians $\hat{h}\left(  Z,j,\zeta\right)  $
are fixed unambiguously, a corresponding total s.a. Dirac Hamiltonian $\hat
{H}\left(  Z\right)  $ is defined in a unique way.

Because s.a. radial Hamiltonians $\hat{h}\left(  Z,j,\zeta\right)  $ are
unique for $Z\leq118$,\ the total Dirac Hamiltonian $\hat{H}\left(  Z\right)
$ with $Z\leq118$ is defined uniquely. For $Z\geq119$, there is a family
$\{\hat{H}_{\nu_{1},...,\nu_{\Delta}}\left(  Z\right)  \}$ of possible\textrm{
}total s.a. Dirac Hamiltonians. The family is parametrized by the parameters
$\nu_{i}\in\mathbb{[}-\pi/2,\pi/2]$, $-\pi/2\sim\pi/2$, $i=1,...,\Delta$. The
number $\Delta$ of the parameters is given by $\Delta=2k(Z)$, where the
integer $k(Z)$ is given by $k(Z)=(1/4+Z^{2}\alpha^{2})^{1/2}-\delta
$,$\;0<\delta\leq1$. Any specific s.a. Dirac Hamiltonian $\hat{H}_{\nu
_{1},...,\nu_{\Delta}}\left(  Z\right)  $ corresponds to a certain
prescription for a behavior of an electron at the origin. The general theory
thus describes all the possibilities that can be offered to a physicist for
his choice. This choice is a completely physical problem. We believe that each
s.a. Dirac Hamiltonian with superstrong\textrm{ }Coulomb field can be
understood through an appropriate regularization of the potential and a
subsequent limit process of removing the regularization. We recall that a
physical interest in the electronic structure of superheavy atoms was
mainly\textrm{ }motivated by a possible pair creation in the superstrong
Coulomb field. Consideration of this effect in the framework of the most
simplest model of a point-like nucleus was accepted to be impossible due to
the conclusion (which is wrong as it is clear now) that this model is
mathematically inconsistent \cite{ZelPo72}. We believe that a rehabilitation
of the model allows returning to a consideration of the particle creation in
this model providing considerable scope for analytical studies.

\begin{acknowledgement}
Gitman is grateful to the Brazilian foundations FAPESP and CNPq for permanent
support; Tyutin thanks FAPESP and RFBR, grant 11-01-00830; Voronov thanks
RFBR, grant 11-02-00685.
\end{acknowledgement}

\end{document}